\documentclass[amsmath,amssymb,aps,prc,floatfix,reprint]{revtex4-1}
\usepackage{bm} 
\usepackage{graphicx}
\usepackage{subfigure}
\usepackage{dcolumn}
\usepackage{multirow}
\usepackage{color}
\usepackage{hyperref}

\newcommand{\reff}[1]{Eq.~\eqref{#1}}
\newcommand{\da}{^{\dagger}}

\begin{document}
\title{Microscopic theory of the \texorpdfstring{\mbox{$\gamma$-decay}}{gamma-decay} of nuclear giant resonances}
\author{Marco Brenna}
\email{marco.brenna@mi.infn.it}
\affiliation{Dipartimento di Fisica, Universit\`a degli Studi di Milano and
INFN, Sezione di Milano, via Celoria 16, I-20133 Milano, Italy}
\author{Gianluca Col\`o}
\email{gianluca.colo@mi.infn.it}
\affiliation{Dipartimento di Fisica, Universit\`a degli Studi di Milano and
INFN, Sezione di Milano, via Celoria 16, I-20133 Milano, Italy}
\author{Pier Francesco Bortignon}
\email{pierfrancesco.bortignon@mi.infn.it}
\affiliation{Dipartimento di Fisica, Universit\`a degli Studi di Milano and
INFN, Sezione di Milano, via Celoria 16, I-20133 Milano, Italy}
\date{\today}

\begin{abstract}
In the past decades, the \mbox{$\gamma$-decay} of giant resonances
has been studied using phenomenological models. In keeping
with possible future studies performed with exotic beams, 
microscopically-based frameworks should be envisaged. In the
present paper, we introduce a model which is entirely based
on Skyrme effective interactions, and treats the ground-state decay
within the fully self-consistent Random Phase Approximation (RPA) 
and the decay to low-lying states at the lowest order beyond
RPA. The model is applied to $^{208}$Pb and $^{90}$Zr, and the results are 
compared with experimental data.
\end{abstract}

\maketitle

\section{Introduction}

Giant resonances (GRs) have been known for several decades to be the clear
manifestation of the existence of nuclear collective motion. They carry definite
quantum numbers (spatial angular momentum $L$, spin $S$, isospin $T$) and, as a
rule, they exhaust a large fraction of the associated energy-weighted sum rule.
Accordingly, the macroscopic picture of a giant resonance is often thought to be
that of a coherent motion of all nucleons. Although a number of experimental
data and theoretical studies have been cumulated, as reviewed in monographic
volumes \cite{bortignon,harakeh}, the question still exists whether we can
access only the inclusive properties of the GRs (energy and fraction of
energy-weighted sum rule), or more exclusive properties associated with the wave
function of the GR. Ultimately, we can say that we miss an unambiguous
confirmation of the macroscopic picture of this collective motion. 

Giant resonances have a finite lifetime. Being excited by one-body external
fields, they are as a first approximation described by coherent superpositions
of 1 particle-1 hole (1p-1h) configurations. The most probable damping mechanism 
is their coupling to progressively more complicated states of 2p-2h $\ldots$
$n$p-$n$h character (up to the eventual compound nucleus state). The associated
contribution to the total width, the so-called spreading width 
$\Gamma^\downarrow$, is the dominant one. The decay width associated with
the emission of one nucleon in the continuum (escape width, $\Gamma^\uparrow$)
is of some relevance in light nuclei but much less important in heavy nuclei. 
The \mbox{$\gamma$-decay} width $\Gamma_\gamma$ is a small fraction ($\approx$
10$^{-3}$) of the total width. Despite this, the study of the \mbox{$\gamma$-decay} of
GRs has been considered a valuable tool since about 30 years
\cite{Beeneexp,beene2}.

In these works, the fact that \mbox{$\gamma$-decay} can be a sensitive probe of the
excited multipolarity, and that $\gamma$-ejectile coincidence measurements can
improve the extraction of the properties of GRs, has been thoroughly discussed.
Generally speaking, the study of the GR decay products (whether particles or
photons) is probably the only way to shed light on the microscopic properties of
the states. To provide an example different from the standard electric GRs
discussed in Refs.~\cite{Beeneexp,beene2}, we can add that in stable nuclei or
in neutron-rich unstable nuclei some information exist on the so-called
low-lying or ``pygmy'' dipole states. Their nature (collective or
non-collective, isoscalar or isovector, compressional or toroidal) is under
strong debate. For these, as for other states, exclusive decay measurements
would be of paramount importance as they could validate some theoretical
picture. 

In this spirit we present here a consistent study of the \mbox{$\gamma$-decay} of giant
resonances, both to the ground and low-lying excited states, not considering the
compound \mbox{$\gamma$-decay} \cite{BeeneGQR}. In the past, the theoretical study of
the \mbox{$\gamma$-decay} of GRs has been undertaken using frameworks like the Nuclear
Field Theory (NFT) \cite{Bortdec} or the Theory of Finite Fermi Systems
\cite{Spethsmall}. These studies have elucidated the basic physical mechanisms
which explain the small \mbox{$\gamma$-decay} probabilities and have provided results
in quite reasonable agreement with experiment. As we discuss below, in
Ref.~\cite{Bortdec} the quenching mechanisms for the decay of the isoscalar
Giant Quadrupole Resonance (ISGQR) to a low-lying isoscalar states, are clearly
pointed out. However, these studies are based on phenomenological models.

After several decades, self-consistent mean-field (SCMF) or density functional
theory (DFT) based models have been developed, and have reached considerable
success for the overall description of many nuclear properties. Among these
models, we can single out those based, respectively, on the nonrelativistic
Skyrme and Gogny effective interactions or on covariant (or relativistic)
effective Lagrangians. Years ago, some of us developed a microscopic description
of the particle decay of GRs based on the use of Skyrme forces
\cite{colo1,colo2}. It is timely to dispose of a fully microscopic description
of the \mbox{$\gamma$-decay} with Skyrme effective interactions and to assess how
large predictive power it can have, and which limitations show up. A further
motivation is provided by the recent measurements carried out at
the Laboratori Nazionali di Legnaro (LNL) \cite{Nicolini}. 

In this work, we develop such model and we apply it to the \mbox{$\gamma$-decay} of the
ISGQR in $^{208}$Pb and $^{90}$Zr, both to the ground-state and to the low-lying
3$^-$ state.
The outline of the paper is the following. In Section~\ref{form} we present the
formalism that we employ. Section~\ref{results} is devoted to the results we
obtain, also comparing them with available experimental data and with other
theoretical calculations found in the literature. In Section~\ref{concl} we draw
our conclusions and eventually, in the \hyperref[appA]{Appendix\ref*{appA}} we
briefly give a
guideline for the calculation of the perturbative diagrams needed for the decay
of RPA excited states into low-lying collective states, as explained in
Section~\ref{form}.

\section{Formalism}\label{form}

In this Section we discuss our theoretical framework. The transition amplitude
for the emission of a photon of given multipolarity from the nucleus, is
proportional to the matrix element of the electric multipole operator
$Q_{\lambda\mu}$. In the long wavelength limit which is appropriate in our case,
this latter operator takes the form
\begin{eqnarray}
Q_{\lambda\mu} & = & \frac{e}{2}\sum_{i=1}^A\left\{\left[
\left(1-\frac{1}{A}\right)^{\lambda}+(-)^{\lambda}
\frac{2Z-1}{A^{\lambda}}\right ] \right.\nonumber\\
&&\left.-\left[\left(1-\frac{1}{A}\right)^{\lambda}+
\frac{(-)^{\lambda+1}}{A^{\lambda}}\right]\tau_z(i)
\right\}r^{\lambda}_ii^{\lambda} Y_{ \lambda\mu}
\left(\hat{\bm{r}}_i \right) \nonumber \\
& \equiv & 
\frac{1}{2}\sum_{i=1}^A e^{eff}_i
r^{\lambda}_i i^{\lambda} Y_{ \lambda\mu}
\left(\hat{\bm{r}}_i \right).
\label{eq:op}
\end{eqnarray}
In this equation, the expression for the effective charge due to the recoil of
the center of mass of the nucleus has been introduced (see, e.g.,
Ref.~\cite{deShalit}).

The gamma decay width, summed over the magnetic substates of the photon and of
the final nuclear state, is then given by 
\begin{equation}
\label{eq:gammawi}
\Gamma_{\gamma}(E\lambda;i\rightarrow f)=
\frac{8\pi(\lambda+1)}{\lambda[(2\lambda+1)!!]^2}
\bigg(\frac{E}{\hbar
c}\bigg)^{2\lambda+1}B(E\lambda; i\rightarrow f),
\end{equation}
where $E$ is the energy of the transition and the reduced transition probability
$B$ associated with the above operator $Q_{\lambda\mu}$ is
\begin{equation}
B(E\lambda;i\rightarrow f)=\frac{1}{2J_i+1}
|\langle J_f\|Q_{\lambda}\|J_i\rangle|^2.
\label{bel}
\end{equation}

\subsection{The \texorpdfstring{\mbox{$\gamma$-decay}}{gamma-decay} to the ground-state}\label{grdec}
\begin{figure}
\includegraphics[width=0.2\textwidth]{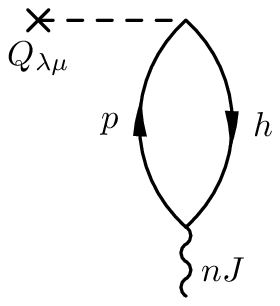}
\includegraphics[width=0.2\textwidth]{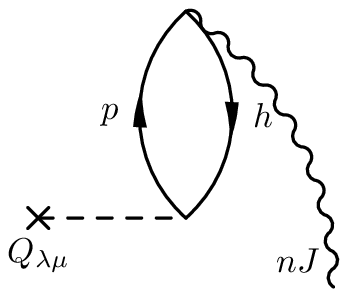}
\caption{Diagrams representing the decay of the vibrational state
$\left|nJ\right\rangle$ state to the ground-state.}
\label{fig:gs}
\end{figure}
We consider in this Subsection the decay of an excited RPA state (that can be,
e.g., a giant resonance) to the ground-state. We limit ourselves to spherical
systems, and the RPA states have quantum numbers $JM$ (we consider natural
parity, or non spin-flip, states for which the orbital angular momentum $L$ is
the same as the total angular momentum $J$); in addition, they are labelled by
an index $n$. Consequently, we can write 
\begin{eqnarray}
&&\left|nJM\rangle \right. = 
O\da_{n}\left(JM\right)\left| RPA \rangle \right., \nonumber\\
&&
O\da_{n}\left(JM\right)
=\sum_{ph}\left[X_{ph}^{nJ}A_{ph}\da\left(JM\right)
-Y_{ph}^{nJ}A_{ph}\left(\widetilde{JM}\right)\right],\nonumber
\label{eq:phon}
\end{eqnarray}
where $O\da_{n}(JM)$ is the creation operator for the state at hand, $\left|RPA \right\rangle$ is the RPA ground-state, $A$ and
$A\da$ are the usual creation
and annihilation operator of a particle-hole (p-h) pair coupled to $JM$, $X$
and $Y$ are the forward and backward RPA amplitudes, and the symbol
$\widetilde{\phantom{J}}$ denotes the time-reversal operation (see, e.g.,
Ref.~\cite{Rowe}).

At the RPA level, in the case of the decay of the state $\left|nJ\rangle\right.$
to the ground-state, we obtain for the reduced matrix of Eq.~\eqref{bel}
with $\lambda$ equal to $J$, 
\begin{equation}
\langle 0 \left\| Q_{J}\right\|nJ\rangle
=\sum_{ph}\left(X_{ph}^{nJ}+Y_{ph}^{nJ}\right) e^{eff}_{ph}
\langle j_p\| i^{\lambda}r^{\lambda}Y_{\lambda}\|j_h\rangle,
\label{eq:RPAdecgs}
\end{equation}
where the effective charge is defined in Eq. (\ref{eq:op}). It is possible to
give a diagrammatic representation of the ground-state decay (see Fig.
\ref{fig:gs}).

\subsection{The \texorpdfstring{\mbox{$\gamma$-decay}}{gamma-decay} to low-lying states}\label{lowdec}

While RPA can be considered an appropriate theory to calculate the ground-state
decay of a vibrational state, the same statement does not hold in the case of a
decay between two vibrational states. The reason is that by construction RPA is
an appropriate theory to describe transition amplitudes between states that
differ only by one vibrational (phonon) state. For other processes, like the one
at hand, the extension to a treatment beyond RPA is mandatory. A consistent
framework which is available is the one provided by the Nuclear Field Theory
(NFT) \cite{NFT1,BortPhysRep}, since this framework takes into account the
interweaving between phonons and single-particle degrees of freedom (or particle-vibration coupling, PVC),
considered
as the relevant independent building blocks of the low-lying spectrum of finite
nuclei. In this work, we consider all the lowest-order contributions to the
\mbox{$\gamma$-decay} between two different phonons. This amounts to writing and
evaluating all lowest-order perturbative diagrams involving single-particle
states and phonon states, that can lead from the initial to the final state by
the action of the external electromagnetic field. The different degrees of
freedom are coupled by particle-vibration vertices. The Nuclear Field Theory, as
mentioned in the introduction, has been already applied to the study of
\mbox{$\gamma$-decay} in Ref.~\cite{Bortdec}. However, the main novelty of the
present
work lies in the consistent use of the microscopic Skyrme interaction. 

\begin{figure*}
\centering
\subfigure[]{\label{fig:graph1}\includegraphics[width=.2\textwidth]{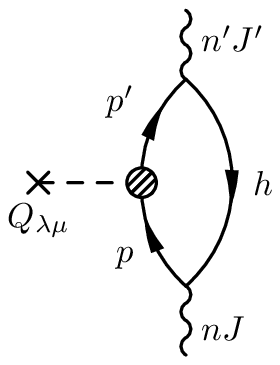}}
\subfigure[]{\label{fig:graph2}\includegraphics[width=.2\textwidth]{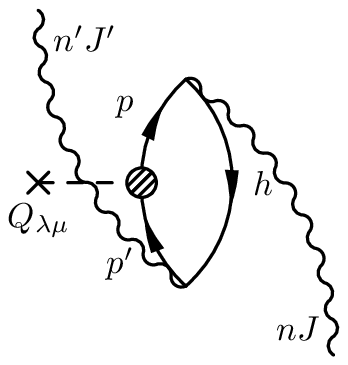}}
\subfigure[]{\label{fig:graph3}\includegraphics[width=.2\textwidth]{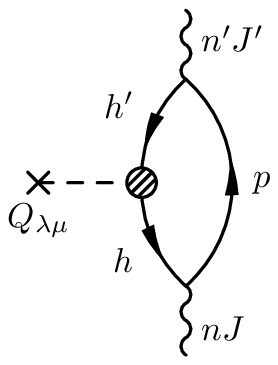}}
\subfigure[]{\label{fig:graph4}\includegraphics[width=.2\textwidth]{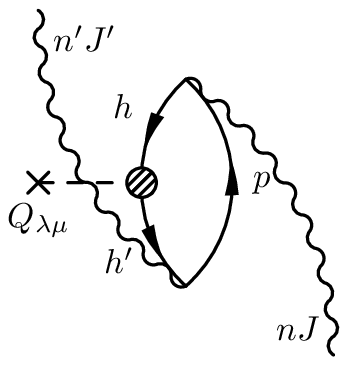}}
\vspace{0.1cm}
\subfigure[]{\label{fig:graph5}\includegraphics[width=.2\textwidth]{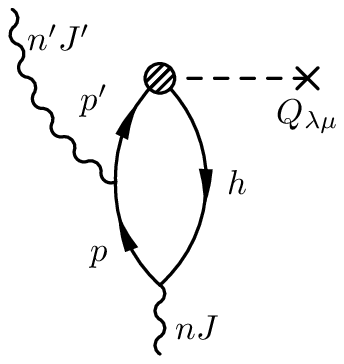}}
\subfigure[]{\label{fig:graph6}\includegraphics[width=.2\textwidth]{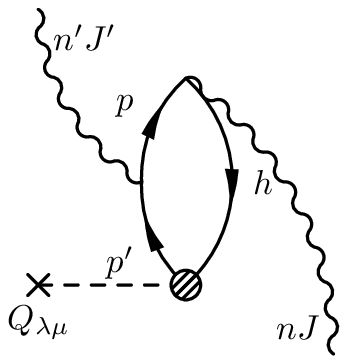}}
\subfigure[]{\label{fig:graph7}\includegraphics[width=.2\textwidth]{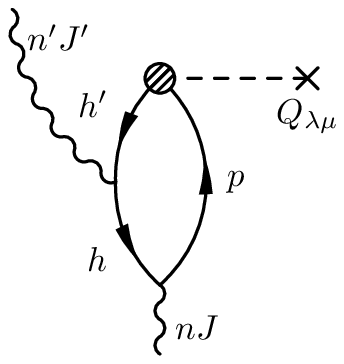}}
\subfigure[]{\label{fig:graph8}\includegraphics[width=.2\textwidth]{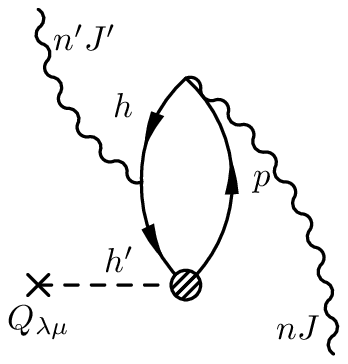}}
\vspace{0.1cm}
\subfigure[]{\label{fig:graph9}\includegraphics[width=.2\textwidth]{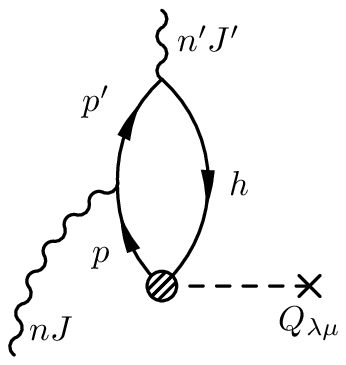}}
\subfigure[]{\label{fig:graph10}\includegraphics[width=.2\textwidth]{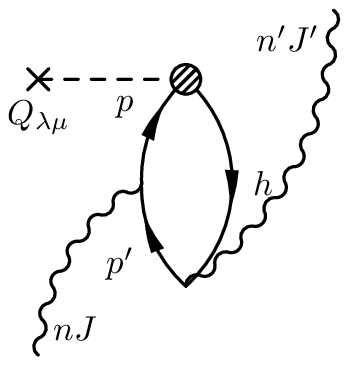}}
\subfigure[]{\label{fig:graph11}\includegraphics[width=.2\textwidth]{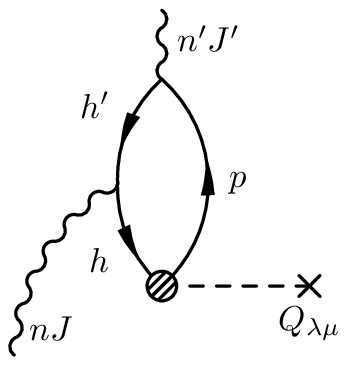}}
\subfigure[]{\label{fig:graph12}\includegraphics[width=.2\textwidth]{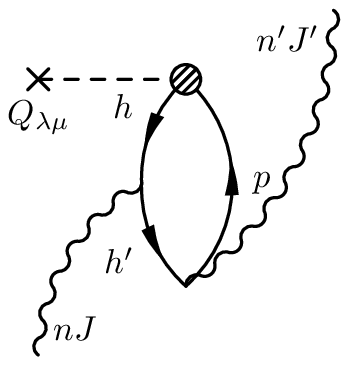}}
{\caption{NFT diagrams contributing to the decay of the $\left|nJ\right\rangle$
state to the $\left|n'J'\right\rangle$ state.}
\label{fig:graphs}}
\end{figure*}

\begin{figure*}[th]
\centering
\includegraphics[width=.9\textwidth]{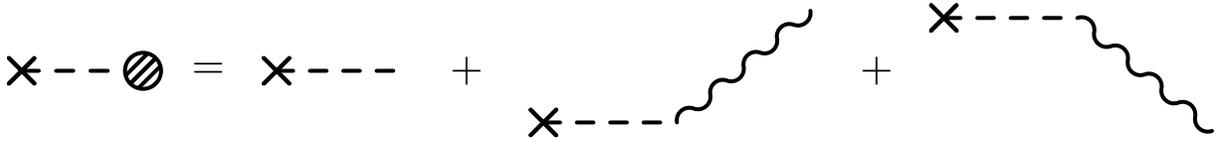}
\caption{Polarization contribution to the operator $Q_{\lambda\mu}$}
\label{fig:pollab}
\end{figure*}

The perturbative diagrams associated with the $\lambda$-pole decay of the
intitial RPA state $\vert n J\rangle$ (at energy $E_J$) to the final state
$\vert n' J' \rangle$ (at energy $E_{J'}$) are shown in Fig.~\ref{fig:graphs},
and the way to evaluate them is sketched in the \hyperref[appA]{Appendix\ref*{appA}}. The resulting
analytic expressions read
{\allowdisplaybreaks
\begin{widetext}
\begin{subequations}
\renewcommand{\theequation}{\theparentequation.\textsc{\alph{equation}}}
\label{eq:dec3-}
\begin{eqnarray}
&&\langle n'J' \left\|
Q_{\lambda}\right\|nJ\rangle_{\text{\subref{fig:graph1}}}=\sum_{pp'h}
(-)^{J+\lambda+J'+1}
\begin{Bmatrix}
J & \lambda & J' \\
j_{p'} & j_h & j_p
\end{Bmatrix}
\frac{\langle p \| V\|h,nJ\rangle\langle h,n'J' \| V\|p'\rangle
Q_{p'p}^{\lambda
pol}}{\left(E_J-\epsilon_{ph}+i\eta\right)\left(E_{J'}-\epsilon_{p'h}\right)},\\
&&\langle n'J' \left\|
Q_{\lambda}\right\|nJ\rangle_{\text{\subref{fig:graph2}}}=\sum_{pp'h}(-)
\begin{Bmatrix}
J   & \lambda & J'\\
j_{p'} & j_h & j_p \\
\end{Bmatrix}
\frac{\langle h \| V\|p,nJ\rangle\langle p',n'J' \| V\|h\rangle
Q_{pp'}^{\lambda
pol}}{\left(E_J+\epsilon_{ph}+i\eta\right)\left(E_{J'}+\epsilon_{p'h}\right)},\\
&&\langle n'J' \left\|
Q_{\lambda}\right\|nJ\rangle_{\text{\subref{fig:graph3}}}=\sum_{hh'p}
\begin{Bmatrix}
J      & \lambda & J'\\
j_{h'} & j_p     & j_h \\
\end{Bmatrix}
\frac{\langle p \| V\|h,nJ\rangle\langle h',n'J' \| V\|p\rangle
Q_{hh'}^{\lambda
pol}}{\left(E_J-\epsilon_{ph}+i\eta\right)\left(E_{J'}-\epsilon_{ph'}\right)},\\
&&\langle n'J' \left\|
Q_{\lambda}\right\|nJ\rangle_{\text{\subref{fig:graph4}}}=\sum_{hh'p}(-)^{
J+\lambda+J'}
\begin{Bmatrix}
J      & \lambda & J'\\
j_{h'} & j_p     & j_h \\
\end{Bmatrix}
\frac{\langle h \| V\|p,nJ\rangle\langle p,n'J' \| V\|h'\rangle
Q_{h'h}^{\lambda
pol}}{\left(E_J+\epsilon_{ph}+i\eta\right)\left(E_{J'}+\epsilon_{ph'}\right)},\\
&&\langle n'J' \left\|
Q_{\lambda}\right\|nJ\rangle_{\text{\subref{fig:graph5}}}= \sum_{pp'h}(-)
\begin{Bmatrix}
 J     & \lambda & J'  \\
j_{p'} &   j_p   & j_h \\
\end{Bmatrix}
\frac{\langle p \| V\|h,nJ\rangle \langle p',n'J' \| V\|p\rangle
Q_{hp'}^{\lambda
pol}}{\left(E_J-\epsilon_{ph}+i\eta\right)\left(E_J-E_{J'}-\epsilon_{p'h}
+i\eta'\right)},\label{eq:graph5}\\
&&\langle n'J' \left\|
Q_{\lambda}\right\|nJ\rangle_{\text{\subref{fig:graph6}}}=\sum_{pp'h}(-)^{
J+\lambda+J'+1}
\begin{Bmatrix}
  J  & \lambda &  J' \\
 j_{p'} & j_p  & j_h \\
\end{Bmatrix}
\frac{\langle h \| V\|p,nJ\rangle \langle p,n'J' \| V\|p'\rangle
Q_{p'h}^{\lambda
pol}}{\left(E_J+\epsilon_{ph}-i\eta\right)
\left(E_J-E_{J'}+\epsilon_{p'h}-i\eta'\right)},\\
&&\langle n'J' \left\|
Q_{\lambda}\right\|nJ\rangle_{\text{\subref{fig:graph7}}}=\sum_{hh'p}(-)^{
J+\lambda+J'}
\begin{Bmatrix}
   J    & \lambda &  J' \\
 j_{h'} &   j_h   & j_p \\
\end{Bmatrix}
\frac{\langle p \| V\|h,nJ\rangle \langle h,n'J' \| V\|h'\rangle
Q_{h'p}^{\lambda
pol}}{\left(E_J-\epsilon_{ph}+i\eta\right)
\left(E_J-\epsilon_{ph'}-E_{J'}+i\eta'\right)},\\
&&\langle n'J' \left\|
Q_{\lambda}\right\|nJ\rangle_{\text{\subref{fig:graph8}}}=\sum_{hh'p}
\begin{Bmatrix}
 J   & \lambda & J'\\
 j_{h'} & j_h  & j_p\\
\end{Bmatrix}
\frac{\langle h \| V\|p,nJ\rangle \langle h',n'J' \| V\|h\rangle
Q_{ph'}^{\lambda
pol}}{\left(E_J+\epsilon_{ph}-i\eta\right)
\left(E_J+\epsilon_{ph'}-E_{J'}-i\eta'\right)},\\
&&\langle n'J' \left\|
Q_{\lambda}\right\|nJ\rangle_{\text{\subref{fig:graph9}}}=\sum_{pp'h}
\begin{Bmatrix}
 J  & \lambda &  J' \\
 j_h & j_{p'}  & j_p \\
\end{Bmatrix}
\frac{\langle p' \| V\|p,nJ\rangle\langle h,n'J' \| V\|p'\rangle
Q_{ph}^{\lambda
pol}}{\left(E_{J'}-\epsilon_{p'h}\right)
\left(E_J+\epsilon_{ph}-E_{J'}+i\eta\right)},\\
&&\langle n'J' \left\|
Q_{\lambda}\right\|nJ\rangle_{\text{\subref{fig:graph10}}}=\sum_{pp'h}(-)^{
J+\lambda+J'}
\begin{Bmatrix}
  J  & \lambda &   J'   \\
 j_h &   j_{p'}   & j_p \\
\end{Bmatrix}
\frac{\langle p \| V\|p',nJ\rangle\langle p',n'J' \| V\|h\rangle
Q_{hp}^{\lambda
pol}}{\left(E_J-\epsilon_{ph}-E_{J'}+i\eta\right)\left(E_{J'}
+\epsilon_{p'h}\right)},\\
&&\langle n'J' \left\|
Q_{\lambda}\right\|nJ\rangle_{\text{\subref{fig:graph11}}}=\sum_{hh'p}(-)^{
J+\lambda+J'+1}
\begin{Bmatrix}
  J   & \lambda & J' \\
 j_p & j_{h'}  & j_h \\
\end{Bmatrix}
\frac{\langle h \| V\|h',nJ\rangle\langle h',n'J' \| V\|p\rangle
Q_{ph}^{\lambda
pol}}{\left(E_J+\epsilon_{ph}-E_{J'}+i\eta\right)
\left(E_{J'}-\epsilon_{ph'}\right)},\\
&&\langle n'J' \left\|
Q_{\lambda}\right\|nJ\rangle_{\text{\subref{fig:graph12}}}=\sum_{hh'p}(-)
\begin{Bmatrix}
  J  & \lambda &   J'   \\
 j_p &   j_{h'}   & j_h \\
\end{Bmatrix}
\frac{\langle h' \| V\|h,nJ\rangle\langle p,n'J' \| V\|h'\rangle
Q_{hp}^{\lambda
pol}}{\left(E_J-\epsilon_{ph}-E_{J'}
+i\eta'\right)\left(E_{J'}+\epsilon_{ph'}\right)}.
\end{eqnarray}
\end{subequations}
\end{widetext}}
In these equations $\epsilon_{ph}$ is equal to the difference of the
Hartree-Fock (HF) single-particle energies $\epsilon_{p}-\epsilon_{h}$, and $V$
is the residual particle-hole interaction: this latter is discussed below,
together with the expression of its reduced matrix elements. In all the energy
denominators we include finite imaginary parts $\eta$ to take into account the
coupling to more complicated configurations not included in the model space.

In all the above equations, the matrix elements of the operator $Q_{\lambda}$
include the contribution from the nuclear polarization (consequently they carry
the label $pol$). They read
\begin{eqnarray}
Q_{ij}^{\lambda pol} &=& \langle i \|Q_{\lambda} \| j \rangle \nonumber\\
&&+\sum_{n'}\frac{1}{\sqrt{2\lambda+1}}\Bigg[\frac{\langle 0\|
Q_{\lambda}\|n'\lambda\rangle\langle i,n'\lambda \|V\|j\rangle}
{(E_J-E_{J'})-E_{n'}+i\eta}\nonumber\\
&&\phantom{+\sum_{n'}}-\frac{\langle
i \| V\|j,n'\lambda\rangle\langle n'\lambda\| Q_{\lambda}\|0\rangle}
{(E_J-E_{J'}
)+E_{n'}+i\eta}\Bigg]\ ,
\label{eq:polNFT}
\end{eqnarray}
where $\vert n' \lambda\rangle$ are the RPA states having multipolarity
$\lambda$ (and lying at energy $E_{n'}$), while the bare operator $Q_{\lambda}$
has been defined in \reff{eq:op}. The polarization contribution, that is, the
second and third term in the latter equation, has the effect of screening
partially the external field. In a diagrammatic way, the bare and the
polarization contributions to Eq. (\ref{eq:polNFT}) are drawn in Fig.
\ref{fig:pollab}.

It should be noted that the diagrams of Fig.~\ref{fig:graphs} are related two by
two by particle-hole conjugation, so that \ref{fig:graphs}\subref{fig:graph1} is
the opposite of \ref{fig:graphs}\subref{fig:graph4} after the substitutions
$h'\rightarrow p'$ and $h \leftrightarrow p$, and the same holds for the pairs
\ref{fig:graphs}\subref{fig:graph2} -- \ref{fig:graphs}\subref{fig:graph3}, 
\ref{fig:graphs}\subref{fig:graph5} -- \ref{fig:graphs}\subref{fig:graph8}, 
\ref{fig:graphs}\subref{fig:graph6} --
\ref{fig:graphs}\subref{fig:graph7}, 
\ref{fig:graphs}\subref{fig:graph9} -- \ref{fig:graphs}\subref{fig:graph12} 
and \ref{fig:graphs}\subref{fig:graph10} --
\ref{fig:graphs}\subref{fig:graph11}. 

As mentioned above, in the present implementation of the formalism we use
consistently different zero-range Skyrme interactions. The single-particle
energies $\epsilon_i$, and the corresponding wavefunctions, come from the
solution of the HF equations. The energies (and $X$ and $Y$ amplitudes) of the
vibrational states are obtained through fully self-consistent RPA
\cite{Colo2007173}. These quantities enter the reduced matrix elements
associated with the PVC vertices. The basic one, that couples the
single-particle state $i$ to the particle-vibration pair $j$ plus $nJ$, is
\begin{eqnarray}
\langle i \| V \| j,nJ\rangle&=&\sqrt{2J+1}
\sum_{ph}X_{ph}^{nJ}V_J(ihjp) \nonumber\\
&&+(-)^{j_h-j_p+J}Y_{ph}^{nJ}V_J(ipjh).
\label{eq:redV}
\end{eqnarray}
$V_J$ is the particle-hole coupled matrix element
\begin{eqnarray}
V_J(ihjp)&=&\sum_{\{m\}}(-)^{j_j-m_j+j_h-m_h}
\langle j_i m_i j_j -m_j|JM \rangle \nonumber\\
&&\times\langle j_p m_p j_h -m_h|JM \rangle 
v_{ihjp},
\label{eq:coupl}
\end{eqnarray}
while $v_{ihjp}$ stands for $\langle j_i m_i j_h m_h|V| j_j m_j
j_p m_p \rangle$. For the detailed derivation of the the reduced matrix element
of \reff{eq:redV} we refer to the Appendix of Ref. \cite{Colo}. In the
\hyperref[appA]{Appendix\ref*{appA}} of the
present paper we discuss the relationships between the reduced
matrix element of \reff{eq:redV} and the other matrix elements that enter the
previous formulas. In our implementation, $V$ used at the PVC vertex includes
the $t_0$,$t_1$,$t_2$ and $t_3$ terms of the Skyrme force.

\section{Results} \label{results}

In this Section, the results obtained from our numerical calculations in
$^{208}$Pb and $^{90}$Zr are discussed. In particular, we focus on the
\mbox{$\gamma$-decay} width 
$\Gamma_{\gamma}$ associated with the decay of the isoscalar giant quadrupole
resonance (ISGQR) either to the ground-state or to the $3^-$ low-lying state in
the two systems. We have employed four different Skyrme forces: SLy5
\cite{Cha:SLy5}, SGII \cite{VanGiai:SGII}, SkP \cite{Doba:SkP} and LNS
\cite{CaoLombardo:LNS}.

In all cases, we start by solving the HF equations in a radial mesh that extends
up to 20 fm (for $^{208}$Pb) or 18 fm (for $^{90}$Zr), with a radial step of 0.1
fm. Once the HF solution is found,
the 
RPA equations are solved in the usual matrix  formulation. Vibrations (or
phonons) with multipolarity $L$ ranging from 1 to 3, and with natural parity,
are calculated. The RPA model space consists of all the occupied states, and all
the unoccupied states lying below a cutoff energy $E_C$ equal to 50 MeV
and 40 MeV for $^{208}$Pb and $^{90}$Zr, respectively. The
states at positive energy are obtained by setting the system in a box, 
that is, the continuum is discretized. These states have increasing values of
the radial quantum number $n$, and are calculated for those values of $l$ and
$j$ that are allowed by selection rules. With this choice of the model space the
energy-weighted sum rules (EWSRs) satisfy the double commutator values at the
level of about 99\%; moreover, the energy and the fraction of EWSR of the states
which are relevant for the following discussion are well converged. 

\begin{table}
\caption{Energy $E$ of the ISGQR and \mbox{$\gamma$-decay} 
width associated with its transition to the ground-state. The first four rows
correspond to the present 
RPA calculations performed with different Skyrme
parameter sets, for the two nuclei at hand. In this case,
for $^{208}$Pb
 we show both the 
bare $\Gamma_\gamma$ from Eq. (\ref{eq:gammawi}) 
as well as the renormalized value which is 
discussed in the main text. The next three rows report the results of previous
theoretical calculations \cite{Bortdec,Spethsmall,BeeneGQR} for $^{208}$Pb. 
In the last row the experimental value 	for $^{208}$Pb
from Ref.~\cite{Beeneexp} is displayed, corresponding to the direct
decay.}\label{tab:decgs}
\begin{ruledtabular}
\begin{tabular}{cccccc}
&\multicolumn{3}{c}{$^{208}$Pb} &  \multicolumn{2}{c}{$^{90}$Zr} \\
& $E$ [MeV] & $\Gamma_{\gamma}$ [eV] &
$\Gamma^{ren}_{\gamma}$ [eV] & $E$ [MeV] &
$\Gamma_{\gamma}$ [eV] \\
\hline
SLy5 & 12.28 & 231.54 & 127.58 & 15.33 & 211.77 \\
SGII & 11.72 & 163.22 & 113.57 & 14.90 & 182.03 \\
SkP  & 10.28 & 119.18 & 159.72 & 13.09 & 107.27 \\
LNS  & 12.10 & 176.57 & 104.74 & 15.48 & 182.71 \\
\hline
Ref. \cite{BeeneGQR}   & 11.20 & \multicolumn{2}{c}{175} & &--\\
Ref. \cite{Bortdec}    & 11.20 & \multicolumn{2}{c}{142} & &--\\
Ref. \cite{Spethsmall} & 10.60 & \multicolumn{2}{c}{112} & &--\\
\hline
Ref. \cite{Beeneexp}   & 10.60 & \multicolumn{2}{c}{130$\pm$40} & &--\\
\end{tabular}
\end{ruledtabular}
\end{table}

\subsection{Ground-state decay}

We group in Table \ref{tab:decgs} the results obtained for the decay of the
ISGQR to the ground-state. In general, our calculations reproduce the experiment
quite well, without any parameter adjustment. They tend at the same time to
overestimate the decay width, and this is true in particular for SLy5; however,
even in this worst case, our result lies within $2.5\sigma$ from the
experimental value. 
\begin{table}
\caption{Detailed properties of the ISGQR and of the low-lying
3$^-$ state in $^{208}\mathrm{Pb}$. The label Exp. indicates the
corresponding experimental values (the italic number after the value is the
experimental error), these values are from Ref.~\cite{Beeneexp} for the ISGQR and from Ref.~\cite{NDS208}
for the 3$^{-}$ state.}
\label{tab:collstates}
\begin{ruledtabular}
\begin{tabular}{c*{4}{c}}
& \multicolumn{2}{c}{$2^+$} & \multicolumn{2}{c}{$3^-$} \\
& E [MeV] & EWSR $[\%]$ & E [MeV] & B(E3$\uparrow$) 
[10$^{5}$e$^2$fm$^{6}$] \\
\hline
Exp. & 10.6  & 90 \textit{20} & 2.6145 \textit{3} & 6.11 \textit{9}\\
SLy5 & 12.28 & 69.27 & 3.62 & 6.54 \\
SGII & 11.72 & 72.31 & 3.14 & 6.58 \\
SkP  & 10.28 & 81.79 & 3.29 & 5.11 \\
LNS  & 12.10 & 66.98 & 3.19 & 5.67 \\
\end{tabular}
\end{ruledtabular}
\end{table}

More importantly, this discrepancy is entirely due to the fact that the
properties of the giant resonance (energy and fraction of EWSR) do not fit
accurately the experimental findings. In particular, the critical quantity turns
out to be the resonance energy, since in \reff{eq:gammawi} the energy of the
transition is raised to the fifth power: consequently, an increase of the energy
by 1 MeV produces an increase of the gamma decay width by about 50\% (at 10
MeV). To substantiate this point, in the last column of Table \ref{tab:decgs} we
report the values obtained for the decay width after having rescaled the ISGQR
energy to the experimental value (shown in Table~\ref{tab:collstates}). In particular, in
Table~\ref{tab:collstates}, the fraction of EWSR exhausted by the ISGQR is shown. The experimental value reported here is from
Ref.~\cite{Beeneexp} and is obtained from the measurement of the \mbox{$\gamma$-decay} to the ground state. On the other hand,
to give an idea of the experimental uncertainty on this observable, we can say that in Ref.~\cite{Beeneexp}, from direct
measurements, a value ranging from 78\% to 98\% is found, depending on the background subtraction, and moreover in the
literature several results in the range $70\%-170\%$ can be found (see e.g. Ref.~\cite{harakeh} or \cite{NDS208}). This is
an indication of systematic uncertainities that include those on the optical potentials used in the experimental analysis.


We can conclude that, since for all the interactions the experimental value of
the
ground-state decay width can be obtained simply by scaling the energy to the
experimental
value, it means that this kind of measurement is not particularly
able to discriminate between models more than usual integral properties.

For completeness, in Table~\ref{tab:decgs} the previous theoretical values
found in literature \cite{Bortdec,Spethsmall,BeeneGQR} are listed as well. In
Ref.~\cite{Bortdec}, the surface coupling model (cf. Ref.~\cite{Bortignon:GR})
is used in order to evaluate the reduced transition probability and the decay
width. In Ref. \cite{Spethsmall}, the Theory of Finite Fermi Systems (cf.
Ref.~\cite{Spethbig}) is implemented with a separable interaction to obtain the
decay width. In Ref.~\cite{BeeneGQR}, finally, the value is estimated from the
empirical energies and fraction of EWSR.

\subsection{The quadrupole strength function}

Before we apply our beyond RPA model to a detailed and exclusive observable such
as the decay from the ISGQR to the 3$^-$ state, it is important to test that, at
the same level of approximation, one can reproduce more general quantities like
the strength function of the ISGQR. It has been known for several decades that
coupling with low-lying vibrations is the main source of the giant resonance
width \cite{BBB}. In Ref. \cite{Bortignon:GR}, calculations of the giant
resonance strength function that take into account this coupling have been
performed, based on the use of a phenomenological separable force in the
surface coupling model. We perform a
similar calculation here by using consistently the Skyrme force SLy5, as
discussed above.

The probability of finding the ISGQR state per unit energy can be written as 
\begin{equation}
P(E)=\frac{1}{2\pi}\frac{\Gamma_\textup{GQR}+\eta}{\left(E-E_\textup{GQR}
-\Delta
E_\textup{GQR}\right)^2+\left(\frac{\Gamma_\textup{GQR}+\eta}{2}\right)^2}
\label{eq:prob}
\end{equation}
where $\Delta E_\textup{GQR}$ is the real part of the sum of the eight
contributions in
Eq.~(\ref{eq:str}), while $\Gamma_\textup{GQR}$ is the imaginary part of the
same sum. The
parameter $\eta$ corresponds to the energy interval over which averages are
taken and represents, in an approximate way, the coupling of the intermediate
states to more complicated configurations. In our calculation we set this
parameter at 1 MeV. The diagrams \ref{fig:str1} -- \subref{fig:str6}
correspond to the self-energy of the particle (or the hole), i.e., the processes
in which the particle or the hole reabsorbs the intermediate excitation
$\lambda$, while diagrams \ref{fig:str3} -- \subref{fig:str8} are vertex
corrections which describe the process in which the phonon is exchanged between
particle and hole.
If $J$ or $\lambda$ is a density oscillation, the latter contributions have
opposite sign with respect to the former one, regardless of the spin and isospin
character of $\lambda$ or $J$, respectively \cite{BBB}.

The eight diagrams \ref{fig:str1} -- \subref{fig:str8} are evaluated by the
following expressions:
\begin{figure*}
\centering
\subfigure[]{\label{fig:str1}\includegraphics[width=.2\textwidth]{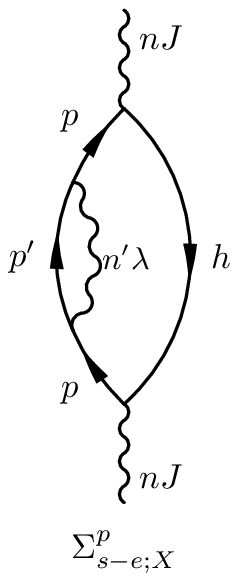}}
\subfigure[]{\label{fig:str5}\includegraphics[width=.2\textwidth]{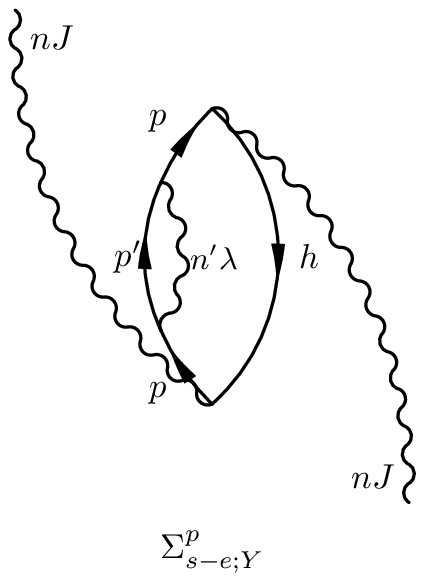}}
\subfigure[]{\label{fig:str2}\includegraphics[width=.2\textwidth]{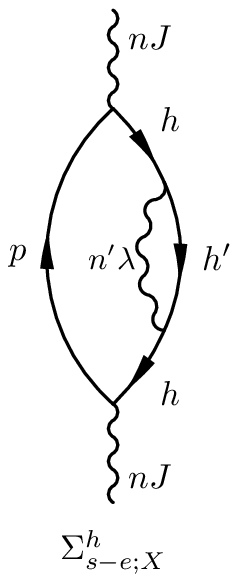}}
\subfigure[]{\label{fig:str6}\includegraphics[width=.2\textwidth]{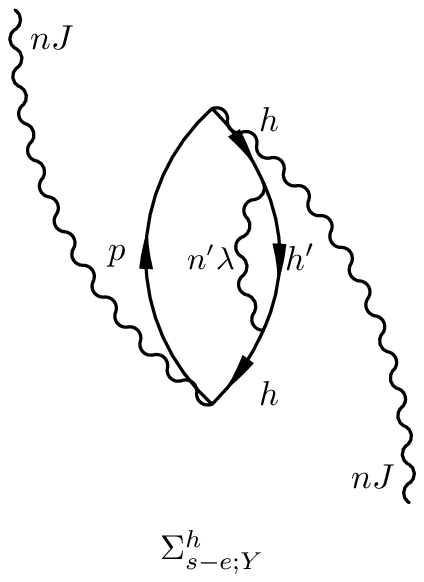}}
\vspace{0.1cm}
\subfigure[]{\label{fig:str3}\includegraphics[width=.2\textwidth]{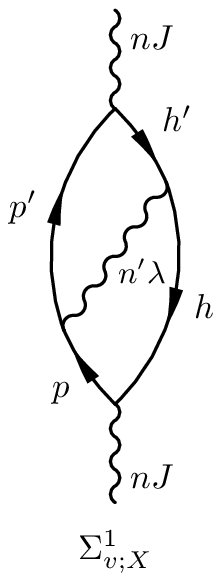}}
\subfigure[]{\label{fig:str7}\includegraphics[width=.2\textwidth]{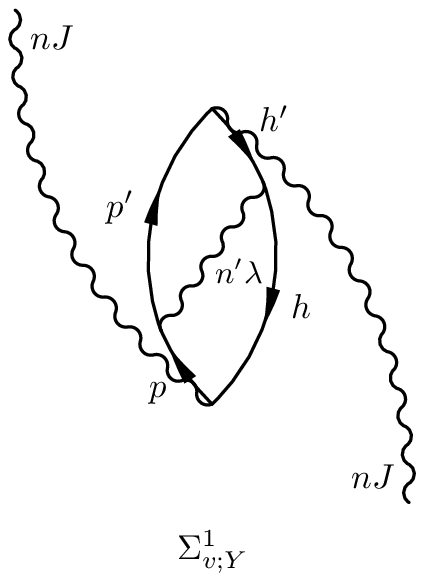}}
\subfigure[]{\label{fig:str4}\includegraphics[width=.2\textwidth]{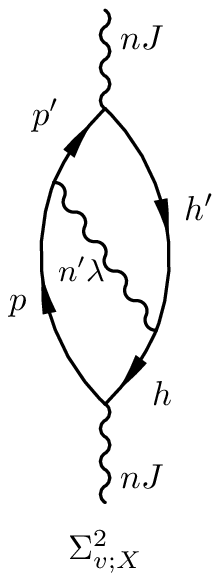}}
\subfigure[]{\label{fig:str8}\includegraphics[width=.2\textwidth]{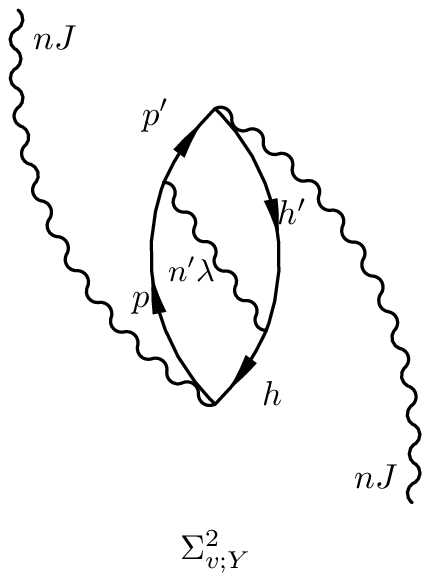}}
{\caption{NFT diagrams contributing to the strength function 
of the giant resonance.}
\label{fig:strgraphs}}
\end{figure*}

{\allowdisplaybreaks
\begin{widetext}
\begin{subequations}
\renewcommand{\theequation}{\theparentequation.\textsc{\alph{equation}}}
\label{eq:str}
\begin{eqnarray}
\Sigma^p_{s-e; X}(GR,E_J)&=&\sum_{pp'hn'}
\frac{1}{(2J+1)(2\lambda+1)}\frac{\left|\langle p \|
V\|h,nJ\rangle\right|^2\left|\langle p \|
V\|p',n'\lambda\rangle\right|^2}{\left(E_j-\epsilon_{ph}
+i\eta\right)^2\left(E_J-E_{n'}-\epsilon_{p'h}+i\eta'\right)}
\\
\Sigma^p_{s-e; Y}(GR,E_J)&=&\sum_{pp'hn'}
\frac{-1}{(2J+1)(2\lambda+1)}\frac{\left|\langle h \|
V\|p,nJ\rangle\right|^2\left|\langle p \|
V\|p',n'\lambda\rangle\right|^2}{\left(E_j+\epsilon_{ph}
+i\eta\right)^2\left(E_J+E_{n'}+\epsilon_{p'h}+i\eta'\right)}
\\
\Sigma^h_{s-e;X}(GR,E_J)&=&\sum_{phh'n'}
\frac{1}{(2J+1)(2\lambda+1)}\frac{\left|\langle p \|
V\|h,nJ\rangle\right|^2\left|\langle h' \|
V\|h,n'\lambda\rangle\right|^2}{\left(E_j-\epsilon_{ph}
+i\eta\right)^2\left(E_J-E_{n'}-\epsilon_{ph'}+i\eta'\right)}
\\
\Sigma^h_{s-e;Y}(GR,E_J)&=&\sum_{phh'n'}
\frac{-1}{(2J+1)(2\lambda+1)}\frac{\left|\langle h \|
V\|p,nJ\rangle\right|^2\left|\langle h' \|
V\|h,n'\lambda\rangle\right|^2}{\left(E_j+\epsilon_{ph}
+i\eta\right)^2\left(E_J+E_{n'}+\epsilon_{ph'}+i\eta'\right)}
\\
\Sigma^1_{v;X}(GR,E_J)&=&\sum_{pp'hh'n'}\frac{(-)^{j_p+j_h+j_{p'}+j_{h'}}}{2J+1}
\begin{Bmatrix}
j_h    & j_p    & J \\
j_{p'} & j_{h'} & \lambda \\
\end{Bmatrix} \nonumber\\
&&\frac{\langle p \| V\|h,nJ\rangle \langle h',nJ \| V \| p'\rangle \langle h
\|V\|h',n'\lambda\rangle\langle p',n'\lambda
\|V\|p\rangle}{\left(E_J-\epsilon_{ph}+i\eta\right)\left(E_J-\epsilon_{p'h'}
+i\eta\right)\left(E_J-E_{n'}-\epsilon_{p'h}+i\eta'\right)}
\\
\nonumber \\
\Sigma^1_{v;Y}(GR,E_J)&=&\sum_{pp'hh'n'}\frac{(-)^{j_p+j_h+j_{p'}-j_{h'
} } } { 2J+1 }
\begin{Bmatrix}
j_h    & j_p    & J \\
j_{p'} & j_{h'} & \lambda \\
\end{Bmatrix} \nonumber \\
&&\frac{\langle h \| V\|p,nJ\rangle \langle p',nJ \| V \| h'\rangle \langle h
\|V\|h',n'\lambda\rangle\langle p',n'\lambda
\|V\|p\rangle}{\left(E_J+\epsilon_{ph}+i\eta\right)\left(E_J+\epsilon_{p'h'}
+i\eta\right)\left(E_J+E_{n'}+\epsilon_{p'h}+i\eta'\right)}
\\
\Sigma^2_{v;X}(GR,E_J)&=&\sum_{pp'hh'n'}\frac{(-)^{j_p+j_h+j_{p'}+j_{h'}}}{2J+1}
\begin{Bmatrix}
j_h    & j_p    & J \\
j_{p'} & j_{h'} & \lambda \\
\end{Bmatrix} \nonumber \\
&&\frac{\langle p \| V\|h,nJ\rangle \langle h',nJ \| V \| p'\rangle\langle  p'
\|V\|p,n'\lambda\rangle\langle h,n'\lambda
\|V\|h'\rangle}{\left(E_J-\epsilon_{ph}+i\eta\right)\left(E_J-\epsilon_{p'h'}
+i\eta\right)\left(E_J-E_{n'}-\epsilon_{ph'}+i\eta'\right)}
\\
\Sigma^2_{v;Y}(GR,E_J)&=&\sum_{pp'hh'n'}\frac{(-)^{j_p+j_h+j_{p'}-j_{h'}}}{2J+1}
\begin{Bmatrix}
j_h    & j_p    & J \\
j_{p'} & j_{h'} & \lambda \\
\end{Bmatrix} \nonumber \\
&&\frac{\langle h \| V\|p,nJ\rangle \langle p',nJ \| V \| h'\rangle\langle  p'
\|V\|p,n'\lambda\rangle\langle h,n'\lambda
\|V\|h'\rangle}{\left(E_J+\epsilon_{ph}+i\eta\right)\left(E_J+\epsilon_{p'h'}
+i\eta\right)\left(E_J+E_{n'}+\epsilon_{ph'}+i\eta'\right)}
\end{eqnarray}
\end{subequations}
\end{widetext}}

\begin{figure}
\centering
\includegraphics[width=0.75\columnwidth,angle=-90]{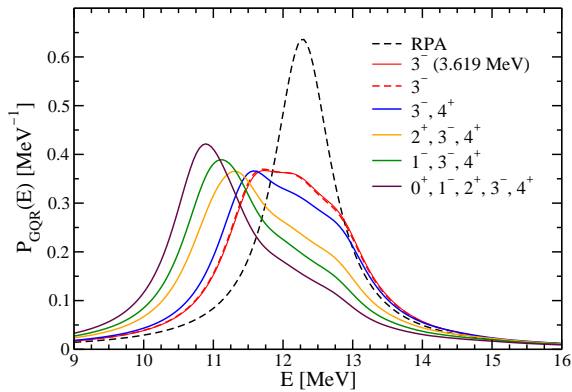}
\caption{(color online) Probability $P$ per unit energy to find the ISGQR at an
energy $E$ in
$^{208}$Pb. Each line corresponds to the probability obtained when the
phonons listed in the legend are used as intermediate states (for the selection
criteria of the phonons we refer to the main text). The red-solid line is the
probability we get when only the low-lying $3^-$ phonon at 3.61 MeV is
considered as intermediate state.
The label \emph{RPA} (black-dashed line) refers to the RPA result, in which none
of the diagrams in Fig.~\ref{fig:strgraphs} are taken into account, but a
lorentzian averaging with functions having 1 MeV width is introduced.}
\label{isgqr_strength}
\end{figure}

The result for the probability of finding the ISGQR, calculated by including in
the diagrams an increasing number of intermediate phonons is displayed in
Fig.~\ref{isgqr_strength}. The RPA model space
is the same used for the computation
of the decay width. Phonons with multipolarity ranging from 0 to 4 and with
natural
parity $(-1)^{\lambda}$, have been considered. Only those having energy smaller 
than 30 MeV and fraction of the total isoscalar and isovector EWSR larger
than 5\% have been selected as intermediate states.
The most important contribution to the spreading width $\Gamma^{\downarrow}$ of
the resonance is given by the low-lying $3^-$ state, while the other phonons do
contribute basically only to the energy shift. We obtain eventually a spreading
width $\Gamma^{\downarrow}$ of the order of 2 MeV, and the energy centroid of
the resonance is shifted down, as compared to the RPA value, to 10.9 MeV. These
results are in good agreement with the experimental findings, that give a
spreading width of~$2.4\pm0.4$ \cite{PhysRevC.22.1832}.



\begin{table}[h]
\caption{Decay width to the low-lying $3^-$ for the interactions 
used, calculated including beyond RPA contributions for the two
nuclei $^{208}$Pb and $^{90}$Zr. 
In particular, for $^{208}$Pb the results from Ref.~\cite{Spethsmall} and
Ref.~\cite{Bortdec} 
are also listed and in the last row, and the experimental value from
Ref.~\cite{Beeneexp} is provided as well.}\label{tab:dec3-}
\begin{ruledtabular}
\begin{tabular}{cc*{2}{D{.}{.}{2.2}}c*{2}{D{.}{.}{2.2}}}
&&\multicolumn{2}{c}{$^{208}$Pb} && \multicolumn{2}{c}{$^{90}$Zr} \\
Interaction &$\;$& \multicolumn{1}{c}{$E_\text{trans}$ [MeV]} &
\multicolumn{1}{c}{$\Gamma_{\gamma}$ [eV]} &
$\;$ &\multicolumn{1}{c}{$E_\text{trans}$ [MeV]} &
\multicolumn{1}{c}{$\Gamma_{\gamma}$ [eV]}\\
\hline
SLy5 && 8.66 & 3.39  && 12.51 & 5.81  \\
SGII && 8.58 & 29.18 && 12.16 & 50.58 \\
SkP  && 6.99 & 8.34  && 10.42 & 5.14  \\
LNS  && 8.90 & 39.87 && 12.72 & 16.95 \\
\hline
Ref.~\cite{Bortdec}   && 8.59 & 3.5 &&& \multicolumn{1}{c}{--}\\
Ref.~\cite{Spethsmall}&& 7.99 & 4   &&& \multicolumn{1}{c}{--}\\
\hline
Ref.~\cite{Beeneexp}  && 7.99 & \multicolumn{1}{c}{5$\pm$5} &&&
\multicolumn{1}{c}{--} \\
\end{tabular}
\end{ruledtabular}
\end{table}

\begin{table}[h]
\caption{The various quenching factors that 
combine to produce the decay width $\Gamma_\gamma$ 
from a typical particle-hole dipole transition, for $^{208}$Pb. 
The decay width reported here refers to a cutoff of 
5\% on the percentage of isovector EWSR for the 
dipole states. The same quantities 
from \cite{Bortdec} are displayed.}
\label{tab:resfin3-}
\begin{ruledtabular}
\begin{tabular}{l*{5}{c}}
& SLy5 & SGII & SkP & LNS & Ref.~\cite{Bortdec}\\
\hline\\
Ph transition [eV] & $10^{3}$ & $10^{3} $& $10^{3}$ & $10^{3}$ & $10^{3}$ \\
\hline\\
 Recoupling coefficient & 3 & 3 & 3 & 3 & 3\\
 $\pi$ -- $\nu$ cancellation  & 5 & 4 & 3--4 & 4 & 4\\
 p -- h cancellation & 3--4 & 2--3 & 2--3 & 3--4 & 2--3\\
 Polarization & 6 & 3 & 7--8 & 4 & 15\\
\hline\\
 $\Gamma_\gamma$ [eV] & 3.39 & 29.18 & 8.34 & 39.87 & 3.50\\
\end{tabular}
\end{ruledtabular}
\end{table}

\subsection{ISGQR decay to the low-lying 3\texorpdfstring{\mbox{$^-$}}{-} state}

In Table~\ref{tab:dec3-} the results obtained for the decay of the ISGQR to the
low-lying octupole state in $^{208}$Pb and $^{90}$Zr are
shown. These correspond to a choice
of the lower cutoff of 5\% on the dipole EWSR of the states considered to
calculate the polarization and a parameter $\eta=2$~MeV
 (for $^{208}$Pb) and
$\eta=3.5$~MeV for $^{90}$Zr;
 these inputs will be clarified and discussed later in the
text.
For completeness, the values found in literature \cite{Spethsmall,Bortdec} for
$^{208}$Pb are listed as well. These are all theoretical results obtained using
different models: in Ref.~\cite{Spethsmall}, the Theory of Finite Fermi Systems
(cf. Ref.~\cite{Spethbig}) with a phenomenological interaction is used to
calculate the decay width, while in Ref.~\cite{Bortdec} the decay width is
obtained by means of the NFT, with a separable interaction at the
particle-vibration vertex.
\begin{table*}
\caption{The effect on the 
\mbox{$\gamma$-decay} width $\Gamma_{\gamma}$ of the width $\Gamma_{D}$ of the
intermetiate dipole states. All the states exhausting the EWSR for more than
5\% are considered. The decay width is an almost
monotonically non-decreasing function of this parameter, as expected from the
Bohr-Mottelson model \cite{BMII}.}
\label{tab:gammaDgamma}
\begin{ruledtabular}
\begin{tabular}{cc*{8}{D{.}{.}{2.2}}}
&$\Gamma_D$ [MeV] & 0.01 & 0.1 & 0.5 & 1.0 & 2.0 & 3.0 & 4.0 & 5.0 \\
\hline
\multirow{4}*{$\Gamma_{\gamma}$ [eV]} 
& SLy5 & 2.34  & 2.35  & 2.46  & 2.69  & 3.39  & 4.32  & 5.34  & 6.36  \\ 
& SGII & 27.28 & 27.19 & 27.00 & 27.25 & 29.18 & 32.68 & 37.21 & 42.25 \\
& SkP  & 7.35  & 7.35  & 7.40  & 7.59  & 8.34  & 9.54  & 11.08 & 12.86 \\
& LNS  & 37.93 & 37.82 & 37.55 & 37.76 & 39.87 & 43.84 & 49.03 & 54.86 \\
\end{tabular}
\end{ruledtabular}
\end{table*}
We discuss here in detail the results we obtain for the decay of
the ISGQR in $^{208}$Pb.
It can be noticed that only two interactions, namely SLy5 and SkP, can
reasonably reproduce the experimental value for the decay width. Nevertheless,
all these forces are able to produce a total
$\Gamma_{\gamma}(\text{ISGQR}\rightarrow 3^-)$ which is only a few percent of
$\Gamma_{\gamma}(\text{ISGQR}\rightarrow \text{g.s.})$, as the experiment
indicates.

In order to understand which of the factors that appear in the several
contributions to the decay width has a major effect on the resulting values, we
have analyzed the sensitivity to the physical inputs in great detail.
Table~\ref{tab:resfin3-} displays, for the four forces used,
the contribution of the several factors included in Eq.~\eqref{eq:dec3-}.
Similar factors from Ref.~\cite{Bortdec} are provided as well. The decay width
that
is obtained considering a typical particle-hole transition is of the order of
$\approx$ keV and can be qualitatively accounted by means of the Weisskopf
estimation
for the reduced transition probability of a single particle
excitation~\cite{BMI}. The label \emph{recoupling coefficient} indicates the
quenching
deriving from the mismatch of angular momenta of the particles involved in the
process. Then, because of the isovector nature of the operator \eqref{eq:op},
diagrams involving protons
and neutrons have opposite sign and partially cancel each other. Moreover,
diagrams in which the operator acts on a particle line must have an opposite
sign to the ones in which it acts on a hole line, reflecting the correlations
between particles and holes in vibrations~\cite{BBB}, resulting in a
compensation of the two contributions. Eventually, the polarization contribution
\eqref{eq:polNFT}, deriving from the screening of the external field by the
mediation of the giant dipole resonance, represent a further and more important
quenching of the original decay width, giving then a final width of the order of
electronvolts.

\begin{figure}
\includegraphics[angle=-90,width=\columnwidth]{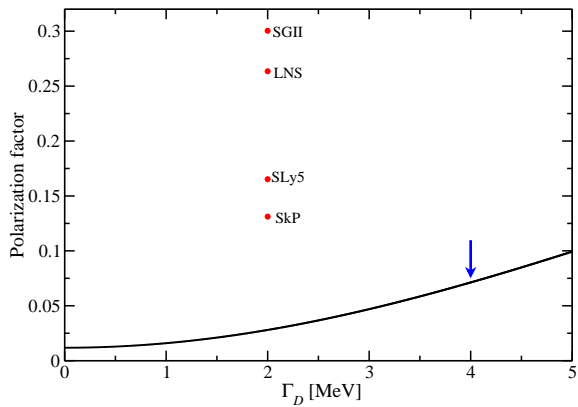}
\caption{(color online) Polarization contribution in the separable framework as
a function of
the parameter $\Gamma_D$ (solid line). The arrow indicates the value used in
Ref.~\cite{Bortdec}. The points are the analogous factors obtained within our
model in which all the dipole states having a fraction of EWSR larger than 5\%
are considered and each one is given a
width $\Gamma_D=2$ MeV.}\label{fig:pol}
\end{figure}

We have studied in particular which assumptions and choices affect the quenching
associated with the polarization
contribution. First of all, in Table~\ref{tab:gammaDgamma}, the variation of the
\mbox{$\gamma$-decay} width $\Gamma_{\gamma}$ with the parameter
$\eta=\frac{\Gamma_D}{2}$ that appears in Eq.~\eqref{eq:polNFT} as imaginary
part of the energy denominator, is discussed. If
only a single dipole
intermediate state is considered, as in Ref.~\cite{Bortdec}, this parameter
should be set equal to the
IVGDR width ($\sim$4 MeV); since in our model, the dipole strength is
fragmented, we should take a smaller value and we
give here the trend of the decay width as a function of this parameter. As
indicated by the plot in Fig.~\ref{fig:pol}, the polarization factor (and
consequently the decay width) should be monotonically non-decreasing when
$\Gamma_D$
increases and reaches a roughly constant value as $\Gamma_D$ goes to zero. In
the
same plot, the points represent the polarization factors that we obtain
using the value 2 MeV for the parameter $\Gamma_D$, but including all the dipole
states having a fraction of EWSR larger than
5\%. 
This value has been chosen in order give a width of the RPA dipole states, each
convoluted with a lorentzian of width equal to $\Gamma_D$, similar to the
experimental IVGDR width. The polarization that we get is then consistent with
the one of Bohr-Mottelson
model~\cite{BMII}, indicated with the arrow in Fig.~\ref{fig:pol}.

We need a lower cutoff on the collectivity of the intermediate states for at
least two reasons: firstly, RPA is known to be not
reliable for non-collective states, and secondly, introducing
them would oblige to take into account the issue of the Pauli
principle correction. We then choose 5\% as lower bound of the isovector and
isoscalar
EWSR, in keeping with several previous works, e.g., Ref.~\cite{Colo}.

A similar analysis carried out on the \mbox{$\gamma$-decay} of the
ISGQR in
$^{90}$Zr into the lowest $3^{-}$ state would bring to analogous
conclusions: the most important effect is the polarization
of the nuclear medium through the excitation of dipole states. Even in this case
the general result is that the decay width to
the octupole state is few percent of the one to the ground-state.
Results from the previously mentioned recent experiment \cite{Nicolini},
are not yet available.

\section{Conclusions}\label{concl}
Our work is motivated by the fact that we deem it is timely to dispose of a fully-microscopic description of some exclusive
properties of giant resonances, like the \mbox{$\gamma$-decay}. In particular, the
\mbox{$\gamma$-decay} has been studied in the past decades using only phenomenological
models. Therefore, we have implemented a scheme in which the single
particle states are obtained within HF, the vibrations are calculated using
fully self-consistent RPA and the whole Skyrme force is employed at the
particle-vibration vertices. 
We treat the ground-state decay within the fully self-consistent RPA and the
decay to low-lying collective vibrations at the lowest contributing order of
perturbation theory beyond RPA.

We have applied our model to the \mbox{$\gamma$-decay} of the isoscalar giant
quadrupole resonance in $^{208}$Pb and $^{90}$Zr into the ground-state and the first low-lying
octupole vibration.
In particular, in $^{208}$Pb, in the case of the ground-state decay, we
find that our outcomes are consistent
with previous theoretical calculations, based on phenomenological models, and
with the experimental data. In particular, all the Skyrme parametrizations give
a \mbox{$\gamma$-decay} width to the ground-state of the order of hundreds of
electronvolts, though, at the same
time, they tend to overestimate it: these discrepancies are due to the fact that
the energy of the resonance does not completely agree with the
experimental data. For this reason, we conclude that the \mbox{$\gamma$-decay} to the
ground-state is not so able to discriminate between different models, at least
not more than other inclusive observable (as energy and strength).

On the other hand, the \mbox{$\gamma$-decay} to low-lying collective states is more
sensitive to the interaction used. As a matter of fact, only two interactions
(namely SLy5 and SkP) manage to achieve a decay width of few electronvolts,
consistently with the experimental finding. In the case of SLy5, this fact is
consistent with the
good features that this parameter set has, as far as
spin-independent processes are concerned (correct
value of the nuclear incompressibility, reasonable
fit of the neutron matter equation of state, good
isovector properties). Nonetheless, the other interactions give a width
$\Gamma_\gamma$ that is of the order of tens of electronvolts and it is very
much quenched with respect to the decay width associated with a single particle
transition. It is quite remarkable that our
calculation, being parameter-free, reproduces
numbers that are several orders of magnitude smaller
than the nuclear scale of $\approx$ MeV. In particular, the description of
the dipole spectrum is a crucial point, because small differences in the
strength of the dipole states, introduced as intermediate states, change
significantly the polarization of the nuclear medium.
For $^{90}$Zr, the general conclusion is similar: the \mbox{$\gamma$-decay} to
low-lying collective states seems to be a
good observable to test the quality of different Skyrme models, being very sensitive to the description of the polarization
of the nuclear medium.

\appendix*

\section{Calculation of the diagrams associated with the decay between
vibrational states}
\label{appA}
\begin{figure*}
\centering
\includegraphics[width=\textwidth]{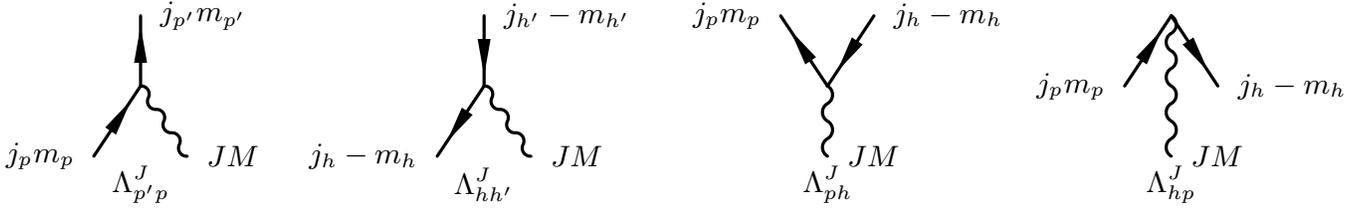}
\caption{Particle-vibration coupling vertices}
\label{fig:PVV}
\end{figure*}

In this Appendix, we provide some details about the calculation of the diagrams
shown in Fig. \ref{fig:graphs}. 

Within the PVC theory, four particle-phonon vertices are possible (cf. Fig.
\ref{fig:PVV}), depending whether the fermionic states involved are particles or
holes. They are related by the particle-hole coniugation operator (cf.
Ref.~\cite{BMII}), so that all the vertices can be brought back to
$\Lambda_{p'p}^J$. From the Appendix of Ref.~\cite{Colo}, we get for the first
vertex $\Lambda_{p'p}^J$ 
\begin{eqnarray}
\Lambda_{p'p}^{J}&=&\langle j_{p'} m_{p'}|V|j_p m_p, nJM \rangle \nonumber\\
&=&(-)^{j_{p'}-m_{p'}}
\begin{pmatrix}
j_{p'} & j_p  & J  \\
m_{p'} & -m_p & -M \\
\end{pmatrix}
\langle p' \|V\|p,nJ\rangle.\nonumber\\
\end{eqnarray}

For example, let us now consider the vertex $\Lambda_{hh'}^J$. We can move the
hole states from the initial to final state (and viceversa) by adding an
appropriate phase factor.
\begin{eqnarray}
\Lambda_{h h'}^{J}&=&\langle (j_{h'} m_{h'})^{-1}|V|(j_h m_h)^{-1}, nJM \rangle
\nonumber \\
&=&(-)^{j_{h'}+m_{h'}+j_h-m_h}\langle j_h
-m_h|V|j_{h'} -m_{h'}, nJM \rangle \nonumber\\
&=&(-)^{j_{h'}-m_{h'}}
\begin{pmatrix}
 j_h  & j_{h'} & J  \\
 -m_h & m_{h'} & -M \\
\end{pmatrix}
\langle h \|V\|h',nJ\rangle \nonumber\\
\label{eq:hh'}
\end{eqnarray}
which is equal to $\Lambda_{p'p}^{J}$ after the identification $h'
\leftrightarrow p$ and $h \leftrightarrow p'$, except for the phase factor.
Similar relations can be established between $\Lambda_{p'p}^{J}$ and
$\Lambda_{ph}^{J}$ or $\Lambda_{hp}^{J}$.
Moreover, the vertices in which the phonon is created instead of annihilated can
be derived from these latter by adding a phase factor $(-)^{J+M}$, changing
the sign
of the projection $M$ of the angular momentum $J$ of the phonon and using the
following expression for the reduced matrix element of the interaction
\begin{eqnarray*}
\langle i,nJ \| V \|j\rangle &=&
\sqrt{2J+1}(-)^{J+j_j-j_i}\sum_{ph}X_{ph}^{nJ}V_J(jhip)\\
&&+(-)^{j_h-j_p+J}Y_{ph}^{nJ}V_J(jpih).
\end{eqnarray*}

These PVC vertices are then used to evaluate the diagrams in
Fig.~\ref{fig:graphs}. In the following, one of them (namely
Fig.~\ref{fig:graphs}\subref{fig:graph5}) is calculated in detail.
For each particle-phonon vertex, we have a reduced matrix element of the
interaction multiplied by a 3-j symbol that takes care of the coupling of
angular momenta. Moreover, the single-particle operator $Q_{\lambda}$ brings
another 3-j symbol and a matrix element. Eventually, the last 3-j symbol,
matching the angular momentum of the initial state, of the final one and of the
operator comes from the Wigner-Eckart theorem, since we need a reduced matrix
element. The energy denominators are obtained by using the rules of second-order
perturbation theory.

{\allowdisplaybreaks
\begin{widetext}
\begin{eqnarray}
\left\langle n'J' \left\|
Q_{\lambda}\right\|nJ\right\rangle_{\ref{fig:graphs}\subref{fig:graph5}}&=&
\sum_{\substack{M\mu m_h\\m_pm_{p'}}}(-)^{J-\lambda+M'}(2J'+1)
\begin{pmatrix}
 J & \lambda & J'\\
 M & \mu     &-M'\\
\end{pmatrix} \nonumber \\
&&\phantom{\sum_{M\mu m_h}}(-)^{J+M}
\begin{pmatrix}
 j_h & j_p &  J  \\
 m_h & m_p & -M \\
\end{pmatrix}
\langle p \| V\|h,nJ\rangle \nonumber \\
&&\phantom{\sum_{M\mu m_h}}(-)^{J'+j_{p'}-m_p}
\begin{pmatrix}
 j_{p'} & j_p & J'  \\
 m_{p'} &-m_p & M' \\
\end{pmatrix}\langle p',n'J' \| V\|p\rangle \nonumber \\
&&\phantom{\sum_{M\mu m_h}}\begin{pmatrix}
 j_h & j_{p'} & \lambda \\
 m_h & m_{p'} & \mu     \\
\end{pmatrix}
\frac{(-)^{j_{p'}-j_h+\lambda}Q_{hp'}^{\lambda pol}}{(E_J-
\epsilon_{ph}+i\eta)(E_J-\hbar \omega_{J'}-\epsilon_{p'h}+i\eta')}
\end{eqnarray}
\end{widetext}}

The four 3-j symbol can be summed in one 6-j symbol by usual relations (see
e.g. Ref.~\cite{BrinkSa})
{\allowdisplaybreaks
\begin{widetext}
\begin{eqnarray}
 \sum_{\substack{M\mu M'\\m_p m_{p'}}}(-)^{j_p+m_p+j_{p'}+m_{p'}+j_h-m_h}&
\begin{pmatrix}
 j_{p'} & j_p & J' \\
-m_{p'} & m_p &-M' 
\end{pmatrix}
\begin{pmatrix}
 j_p & j_h & J \\
-m_p &-m_h & M 
\end{pmatrix}
\begin{pmatrix}
 j_h & j_{p'} & \lambda \\
 m_h & m_{p'} & \mu 
\end{pmatrix}
\begin{pmatrix}
 J & \lambda & J \\
 M & \mu & -M'
\end{pmatrix} \nonumber \\
&=-\frac{1}{\sqrt{2J'+1}}
\begin{Bmatrix}
 J & \lambda & J \\
j_{p'} & j_p & j_h
\end{Bmatrix}
\end{eqnarray}
\end{widetext}}

We then finally get Eq.~\eqref{eq:graph5}

\end{document}